# One- and Multi-Pass Long-Hop Routing for Wireless Network*

Przemysław Błaśkiewicz, Jacek Cichoń, Jakub Lemiesz,
Mirosław Kutyłowski, Marcin Zawada
Wroclaw University of Technology, Wroclaw, Poland
Email: {firstname.lastname}@pwr.edu.pl

## I. INTRODUCTION

Radio is a widely used means of communication for many devices. As such, the devices constitue a radio network, which enables them to pass information from one to another (unicast) or one to many (broadcast). Each radio network operates a given set of frequencies (channels) to transmit, but in majority of settings this set is limited to one channel only. While that brings simplicity to the radio hardware and avoids problems of choosing the right channel by all communicating devices, it also causes the problem of channel occupation.

When two stations transmit at the same time, their communication may reach other stations simultaneously, making it impossible to receive either of them. This is called a *collision*, and there are many MAC (medium-access) protocols designed to avoid that. They can utilise a time-division approach, where each station is given its timeshare when only it is allowed to transmit. Others leverage the hardware capability of carrier sensing – CS (there are also software-based versions of CS), where a station listens to the channel prior to transmitting to assure the channel is free. Some protocols enforce a game on the stations – the one station that wins becomes a *leader* and is allowed access to the channel. This last group is often referred to as *leader election* problems, applicable not only in the realm of radio communication. In this paper we consider a form of randomized leader election algorithm for an efficient routing of a message.

The routing concept is present when information originates at one point of a network and is destined to another. Depending on additional characteristics of the network, many different approaches for routing are implemented. When all stations are in each others' transmission range this problem trivially reduces to a case of the leader election. However, when the message must be passed by intermittent stations (multi-hop network), then means of deciding which stations should be on its way must be devised. In dynamic settings, or when geometry of the network is unknown, this must be done locally by each station involved. The easiest form of information dissipation is flooding, wherein each station, upon receiving the message, re-transmits it blindly. This brings many collisions and unnecessary bandwith use, so other routing methodologies are in use to prevent some stations from re-transmitting the message. In the gradient-based approach ([3], [8]) a virtual map of information spread in the network is constructed and the message is routed towards its highest (for query-type messages) or smallest (for infusion-type) concentration. In the probabilistic approach the next-hop station is chosen from a set of possible stations uniformly (random-walk), or based on probability distribution of some factor available to the re-transmitting station: message history ([5]). Moreover, fundamental question is whether to route over many short hops (short-hop routing) or over a fewer number of longer hops (long-hop routing). In paper [4], the author shows twelve reasons that long-hops could be better than widely supported short-hops.

In our paper we provide mathematical analysis of a probabilistic long-hop routing algorithms which uses as the randomizing factor the estimate of distance of a station from the previous-hop source of the message.

## II. THE MODEL

A communication can be established between transmitter and receiver only when strength of the received radio signal is greater than the receiver antenna sensitivity threshold. The decreasing of signal power between transmitter and receiver is called *path loss*. We use one of the most widely accepted models for network simulations and theoretical analysis. The power received $P$ by a remote station from the sender can be expressed as follows

$$P = P_0 \cdot \left(\frac{d_0}{d}\right)^\alpha \qquad (1)$$

where $P_0$ is the received signal power at distance $d_0$ from a transmitter and $\alpha$ is the path loss exponent.

Since at the receiver side we can measure the signal-to-noise ratio (SNR) then using the equation (1) we estimate the transmitter distance.

## III. ONE-PASS ROUTING

In this section, we are study scenarios of passing a message from station to another station under condition that the message is valid only during some short time interval. Thus, after the time elapsed, we can simply discard this message. Therefore, for us it is important to send this message quickly, but not every message has to reach the destination. It suffice

*Supported by National Centre for Research and Development project no. PBS1/A3/6/2012

if the message achieve the target station in as small number of hops as possible and with reasonable probability.

For described scenerio, we introduce one-pass Algorithm 1 and analyse a single iteration of the process of passing a message. Without loss of generality it can be either of the two cases: (1) some station originated a message and sends it in the current slot, or (2) some station heard a message in the previous slot and won in that slot, hence it is the only station transmitting in the current slot.

Each listening station executes the code given in Algorithm 1, while sending station is starting transmission at will. In the algorithm we use some method of estimating distance from the sending station to the receiver. Admittedly, this does not have an obvious solutions, but some attempts to achieve that goal have been reported.

Once the distance from the transmitting station is obtained, a station uses it to determine if it should pass the message on in the next slot. This decision is made *locally* and independently by each station. What differentiates the outcome of the stations' decisions is their position estimate. Therefore, we analyse the algorithm considering two cases – for random and fixed positions of the stations in the next two subsections, respectively.

---

**Algorithm 1** ONEPASSROUTING

**Setup:**
1: $g(x)$ - special functions analysed in Section III

**Main algorithm:**
1: **if** received message $M$ **then**
2:     $x \leftarrow$ distance estimation based on SNR
3:     **if** rand$(0,1) < g(x)$ **then**
4:        send message $M$ in the next slot
5:     **end if**
6: **end if**

---

*A. Fixed Positions*

To analyse the propagation properties of Algorithm 1 let us assume that nodes are placed on the line and that a sending node is at point 0. Let us also assume that there are $n$ nodes within the range of the sending node and that the range is 1. Therefore, a position of a given node $i$ is a number $x_i \in [0,1]$. Obviously, such normalization will facilitate further analysis but has no influence on the actual properties of the algorithm. The function $g(x)$ in Algorithm 1 expresses the preferences which node should become a leader. For example, setting $g(x) = 1/n$ would give us a classic leader election algorithm, in which all nodes have equal probability to become a leader. If $g(x)$ increases in $x$, the more distant position $x_i$ of a given node, the greater probability that it will become a leader.

We will assess the effectiveness of a given function $g(x)$ by determining the expected length of a hop. Namely, let $S$ be a random variable denoting the number of a winning node and $X_S$ represents the length of a hop, which is a distance between the sending and the winning node. Then we have

$$E[X_S] = \sum_{i=0}^{n} E[X_S|S=i]P[S=i] ,$$

where for $1 \leq i \leq n$

$$x_i = E[X_S|S=i]$$

and

$$P[S=i] = g(x_i) \prod_{j \neq i}(1 - g(x_j)) .$$

Note that we define $S = 0$ when there is no winner and the sending node remains a leader. Since $E[X_S|S=0] = 0$, the first term of the above sum can be always omitted.

Let us now find values of $E[X_S]$ for chosen functions $g(x)$ and some representative arrangements of $n$ nodes on the line. Namely, let us consider three scenarios:

i) uniform distribution of nodes

$$x_i = \frac{i}{n+1} ,$$

ii) unfavourable distribution of nodes

$$x_i = \begin{cases} \dfrac{1}{n+1} & \text{for} \quad 1 \leq i \leq n-1 , \\ \dfrac{n}{n+1} & \text{for} \quad i = n , \end{cases}$$

iii) favourable distribution of nodes

$$x_i = \begin{cases} \dfrac{1}{n+1} & \text{for} \quad i = 1 , \\ \dfrac{n}{n+1} & \text{for} \quad 2 \leq i \leq n . \end{cases}$$

If we set $g(x) = 1/n$ as in classic leader election problem, for presented scenarios we get
i) $E[X_S] = 2/(en) + O(n^{-2})$,
ii) $E[X_S] = 1/(2e) + O(n^{-1})$,
iii) $E[X_S] = 1/e + O(n^{-1})$.

It is clear that an arrangement of nodes is strictly connected to the expected length of the hop in the algorithm. To increase that length more distant nodes should be somehow preferred.

Let us set $g(x) = x^n$ to express our preference of more distant nodes. Then, we can show that for presented scenarios we get
i) $E[X_S] = 1/e + o(1)$,
ii) $E[X_S] = \alpha + o(1)$, $\alpha > 0.4$.
iii) Unfortunately, in this scenario we get $E[X_S] \to 0$ as $n \to \infty$, which results from the fact that most of nodes transmit with large probability.

Conversely, if we choose $g(x) = x/n$, then we get in scenarios
i) $E[X_S] \to 0$ as $n \to \infty$,
iii) $E[X_S] = 2/e + O(n^{-1})$.

The above analysis shows that the right selection of function $g(x)$ might significantly increase the expected length of a hop. On the other hand, analysed cases suggest that it is a nontrivial task to choose function $g(x)$ that would be suitable for an arbitrary distribution of nodes (i.e. without some a priori knowledge concerning their arrangement). In the next section we analyse another interesting scenarios, in which nodes are arranged according to some given probability distribution.

*B. Random Positions*

As we said in the Introduction, we model our network as a line of length $L$ of $n$ nodes. To reflect a random positioning of nodes along this line, we use $X_1, \ldots, X_n$ to represent their positions, s.t. $X_i \sim \mathcal{U}(0, L)$ are independent uniform random variables for $i = 1, \ldots, n$. The distribution of stations (hence: the distribution of $X_i$'s) determines the probability with which each station can win the contest in a slot, so we let $S^1(X_1, \ldots, X_n)$ be a random variable taking values in $\{1, \ldots, n\}$ to represent the station number of the winning station. Consequently, the probability for a given station $i$ to win the contest $P(S^1(X_1, \ldots, X_n) = i)$ is equal to

$$\int_{[0,L]^n} P(S^1(x_1, \ldots, x_n) = i) \cdot f_{(X_1, \ldots, X_n)}(x_1, \ldots, x_n) d(x_1 \ldots x_n),$$

where $f_{(X_1, \ldots, X_n)}$ is the density of the probability distribution of $X_1, \ldots, X_n$.

Let us fix $i$ and focus on probability that $i$th station wins in a slot according to Algorithm 1. The station's position is $x_i$ and it decides to transmit with probability $g(x_i)$, so the probability of winning is:

$$P(S^1(x_1, \ldots, x_n) = i) = g(x_i) \prod_{j \neq i}(1 - g(x_j)). \quad (2)$$

Which means, that $i$-th station transmitted successfully since the remaining stations decided to hold in that round. By Fubini theorem and the assumption that $X_i$'s are independent one can write $f_{(X_1, \ldots, X_n)}(x_1, \ldots, x_n) = f_{X_1}(x_1) \cdot \ldots \cdot f_{X_n}(x_n)$, and the sought probability $P(S^1(X_1, \ldots, X_n) = i)$ is now

$$\int_0^L g(x_i) f_{X_i}(x_i) dx_i \prod_{j \neq i} \int_0^L (1 - g(x_j)) f_{X_j}(x_j) dx_j. \quad (3)$$

Since the goal of the algorithm is to pass the message as far as possible in one hop, we introduce another random variable to determine message's progress. Let $X_{S^1(X_1, \ldots, X_n)}$ be such a random variable – it represents the distance between the station transmitting in the current slot and the station that won in that slot. Maximizing the message progress equals maximizing the expected value of that random variable with respect to the decision function $g(x)$.

First, we proof the following theorem:

**Theorem 1.** *The expected progress $E[X_{S^1(X_1, \ldots, X_n)}]$ of transmitting a message at one round is given by*

$$\sum_{i=1}^n \int_0^L x_i P(S^1(X_1, \ldots, x_i, \ldots, X_n) = i) f_{X_i}(x_i) dx_i$$

*Proof:* Notice that $E[X_{S^1(X_1, \ldots, X_n)}]$ can be expressed as

$$\sum_{i=1}^n E[X_i | S^1(X_1, \ldots, X_n) = i] P(S^1(X_1, \ldots, X_n) = i).$$

Next, we can modify the above equation by expanding the conditional expected value and by assumption that $X_i$ has density function $f_{X_i}$:

$$\sum_{i=1}^n \int_0^L x_i f_{X_i}(x_i | S^1(X_1, \ldots, X_n) = i) dx_i \cdot P(S^1(X_1, \ldots, X_n) = i).$$

Recall that $S^1(X_1, \ldots, X_n)$ is a discrete random variable. Hence, applying Bayes theorem, we further obtain

$$\sum_{i=1}^n \int_0^L x_i \frac{P(S^1(X_1, \ldots, X_n) = i | X_i = x_i) f_{X_i}(x_i)}{P(S^1(X_1, \ldots, X_n) = i)} dx_i \cdot P(S^1(X_1, \ldots, X_n) = i).$$

We get desired formula, after simplification of the above equation. ∎

By the same argument used for equation (3) and Theorem 1, we obtain a closed formula for the expected progress of a message in one round:

$$\sum_{i=1}^n \int_0^L x_i g(x_i) f_{X_i}(x_i) dx_i \prod_{j \neq i} \int_0^L (1 - g(x_j)) f_{X_j}(x_j) dx_j.$$

As we mention, the function $g(x_i)$ introduced in Algorithm 1 determines the expected progress of the message, and Theorem 1 provides us means to determine its optimal form.

Ideally, one should maximize it w.r.t all possible functions $g(x_i)$. However, at this point we are able to find closed formulas for selected functions:

i) Let stations make their decisions independently and uniformly without distance estimation and $X_i \sim \mathcal{U}(0, 1)$. We know (see [6]) that in this case optimal function will be $g(x) = 1/n$. Putting that into Eq. (3) we get

$$P[S^1(X_1, \ldots, X_n) = i] = \frac{1}{n}\left(1 - \frac{1}{n}\right)^{n-1}.$$

Then, by Theorem 1 the expected progress follows:

$$E[X_{S^1(X_1, \ldots, X_n)}] = \frac{1}{2}\left(1 - \frac{1}{n}\right)^{n-1}.$$

ii) On the other hand, let function $g$ depend on position such that stations further away have greater probability of becoming a leader. We keep that $X_i \sim \mathcal{U}(0, 1)$. Based on simulations and ease of calculations, we choose $g(x) = x^{n-1}$. In this case, we can calculate the expected progress as

$$E[X_{S^1(X_1, \ldots, X_n)}] = \frac{n}{n+1}\left(1 - \frac{1}{n}\right)^{n-1}.$$

Clearly, these two cases show us that we can improve the expected message progress from $1/(2e)$ to $n/((n+1)e)$, which means that asymptotically we obtain a double improvement. However, if we choose the same function for all station, then the maximal probability cannot be greater than $1/(n \cdot e)$ in one hop. Therefore, in the next section we introduce also algorithm that solves this issue.

## IV. MULTI-PASS ROUTING

In this section, we extend the one-pass Algorithm 1 to the multi-pass case, namely Algorithm 2. This means that we want to increase the probability of reaching the target station. We would like to achieve this simply i.e. we repeat transmission until a forwarded message was send successfully.

---

**Algorithm 2** MULTIPASSROUTING

**Setup:**
1: $g^m(x)$ - special function defined in the paper

**Main algorithm:**
1: **if** received message $M$ **then**
2:     $x \leftarrow$ distance estimation
3:     $m \leftarrow 0$
4:     **repeat**
5:         **if** $\text{rand}(0,1) < g^m(x)$ **then**
6:             send message $M$ in the next slot
7:         **end if**
8:         $m \leftarrow m+1$
9:     **until** some station transmission was successful
10: **end if**

---

For this generalized algorithm, let us calculate the probability for a particular $i$-th station of becoming a leader, that is the probability $P(S^\infty(x_1,\ldots,x_n) = i)$, where $S^\infty(x_1,\ldots,x_n)$ is random variable denoting the station number which became a leader. As before, the position of the station is $x_i$ and it retransmits with probability $g^m(x_i)$ in $m$-th round. For a general case of $m$-th round leader election, we extend Eq. 2 from the single-round, so that this time the probability $P(S^\infty(x_1,\ldots,x_n) = i)$ is:

$$\sum_{m=0}^{\infty} g^m(x_i) \prod_{j\neq i}(1 - g^m(x_j)) \cdot \left(1 - \sum_k g^m(x_k) \prod_{j\neq k}(1 - g^m(x_j))\right)^m.$$

If we choose the function $g^m(x)$ such that it does not depend on the $m$ i.e for all $m$ the function $g^m(x) = g(x)$. Then the infinite geometric series can be transformed into a more succinct form yields:

$$P(S^\infty(x_1,\ldots,x_n) = i) = \frac{g(x_i)\prod_{j\neq i}(1-g(x_j))}{\sum_{k=1}^{n} g(x_k)\prod_{j\neq k}(1-g(x_j))}.$$

Then, we can get similar theorem to Theorem 1:

**Theorem 2.** $E[X_{S^\infty(X_1,\ldots,X_n)}]$ is equal to

$$\sum_{i=1}^{n} \int_0^L x_i P(S^\infty(X_1,\ldots,x_i,\ldots,X_n) = i) f_{X_i}(x_i) dx_i \quad (4)$$

*Proof:* The proof is omitted. It is similar to the Theorem 1. ∎

Ideally, as in the previous section, one should maximize it w.r.t all possible functions $g(x_i)$. However, at this point we are able to find closed formulas for selected functions:

i) Consider $g^m(x) = p$, meaning that all stations who heard the message transmit with the same probability, regardless of their distance estimation. Then the probability $P(S^1(x_1,\ldots,x_n) = i) = p(1-p)^{n-1}$. Now we calculate the probability $P(S^\infty(x_1,\ldots,x_n) = i)$ as

$$\frac{p(1-p)^{n-1}}{\sum_{k=1}^{n} p(1-p)^{n-1}} = \frac{1}{n}.$$

Again, by setting constant probability of retransmission to all stations we got uniform probability of becoming the leader. Then, assuming that $X_i \sim \mathcal{U}(0,1)$, we can calculate the expected travel distance as

$$E[X_{S^\infty(X_1,\ldots,X_n)}] = \frac{1}{2}.$$

ii) Let consider function defined as

$$g^m(x) = \begin{cases} x^{n-1} & \text{for } m = 0 \\ p & \text{for } m \geq 1 \end{cases}.$$

It means that at first round station use function $x^{n-1}$ and then it follows constant distribution $p$. Then $P(S^\infty(x_1,\ldots,x_n) = i)$ is equal to

$$x_i^{n-1}\prod_{j\neq i}(1-x_j^{n-1}) + \frac{1}{n}\Big(1 - \sum_{k=1}^{n} x_k^{n-1}\prod_{j\neq k}(1-x_j^{n-1})\Big).$$

Then, the expected distance is

$$E[X_{S^\infty(X_1,\ldots,X_n)}] = \frac{1}{2}\Big(1 + \frac{n}{n+1}(1-\frac{1}{n})^n\Big)$$

Asymptotycally it improve the distance from $1/2$ to $1/2 + 1/2e$ at one hop.

Notice that in the scenario (ii) we obtain a hybrid solution i.e. we combine two function from previous section. This solution has two important advantages. Firstly, we get solution that is more stable. As simulation shows it progress more rapidly. Secondly, we are able to find analytical closed form solution.

## V. SIMULATIONS

In order to test the analytic results, we implemented the algorithms in a network simulator [2]. The parameters used for simulation are gathered in Table V.

TABLE I
PARAMETERS SET USED FOR SIMULATION

| $N$ | 20, | 200 | | | | | |
|---|---|---|---|---|---|---|---|
| $d$ | 200, | 300, | 400, | 600, | 1000, | 2500, | 5000, 7500 |
| $m$ | $-5$, | 0, | 1, | 5 | | | |
| $r$ | 130 | | | | | | |

The locations of $N$ nodes along an $d$-long line is uniform and random, which, given theoretical range $r$ of a single station yields the number of stations within its range:

$$n = N/d \cdot r.$$

Importantly, to reflect (slight) variation between stations (eg. resulting from different levels of available power for transmission, etc.), we varied their transmission powers in the range $\pm 10\%$ from the value of $-5$dBm. The pathloss model used for simulation was "IndoorPathLoss" (see [2]). The range was computed by inverting function used for calculating the path loss for microcells as specified in [1] and subsequently employed in [7]:

$$\log_{10}(r) = (L_{max} - 15.3)/37.6,$$

where $L_{max}$ (maximal path loss) is the difference between transmit power and receiver sensitivity.

The value measured in the simulation was the distance covered by a message in a single transmit-receive cycle. We extracted from simulator's run the distances between the station that successfuly transmitted (propagated) a message and the station that sent that message to the propagating station. To adhere to analytical part of this paper, the measured distance was normalized by $r$, i.e. by (theoretically) maximal range over which a message can be transmitted. Notably, due to variations in transmit power and path loss models, that normalized value was occasionally greater than one.

For each combination of parameters from Table V a total of 500 trials were simulated, each with different seed to random generators. The average (normalized) distance was then extracted along with its standard deviation, and plotted as a function of $n$. Results for randomly spaced and equally spaced stations on the line are given in Figures 1.

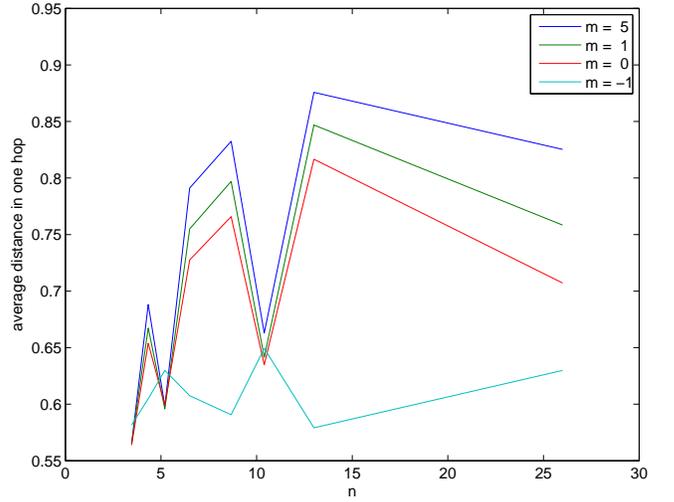

Fig. 1. Average (normalized) distance travelled by message in scenario for randomly spaced stations. X-axis shows estimated number of stations present in the transmission range of each station (station density).

## APPENDIX

We prove an equation from Section IV scenerio (ii)

$$E[X_{S^\infty(X_1,\ldots,X_n)}] = \frac{1}{2}\bigl(1 + \frac{n}{n+1}(1 - \frac{1}{n})^n\bigr)$$

*Proof:* Since $E[X_{S^\infty(X_1,\ldots,X_n)}]$ is given by

$$\sum_{i=1}^{n} \int_0^1 x_i P(S^\infty(X_1,\ldots,x_i,\ldots,X_n) = i) f_{X_i}(x_i) dx_i$$

We need to calculate $P(S^\infty(X_1,\ldots,x_i,\ldots,X_n) = i)$ which can be done from the following equation:

$$\int_0^1 \ldots \int_0^1 P(S^\infty(x_1,\ldots,x_n) = i) \cdot f_{X_1}(x_1) \cdot \ldots \cdot f_{X_{i-1}}(x_{i-1}) \cdot$$
$$f_{X_{i+1}}(x_{i+1}) \cdot \ldots \cdot f_{X_n}(x_n) \, dx_1 \ldots d_{x_{i-1}} d_{x_{i+1}} \ldots d_{x_n}.$$

Since $P(S^\infty(x_1,\ldots,x_n) = i)$ is

$$x_i^{n-1} \prod_{j \neq i}(1 - x_j^{n-1}) + p(1-p)^{n-1} \cdot$$
$$(1 - \sum_{k=1}^{n} x_k^{n-1} \prod_{j \neq k}(1 - x_j^{n-1})) \cdot$$
$$(1 + (1 - np(1-p)^{n-1}) + (1 - np(1-p)^{n-1})^2 + \ldots).$$

Thus, after simplification of the above formula, we obtain

$$x_i^{n-1} \prod_{j \neq i}(1 - x_j^{n-1}) + \frac{1}{n}\bigl(1 - \sum_{k=1}^{n} x_k^{n-1} \prod_{j \neq k}(1 - x_j^{n-1})\bigr).$$

The above equation we can be integrated term-by-term:

i) the first term, we can integrate different variables separately: $\sum_{i=1}^{n} \int_0^1 \ldots \int_0^1 x_i \cdot x_i^{n-1} \prod_{j \neq i}(1 - x_j^{n-1}) f_{X_1}(x_1) \cdot \ldots \cdot f_{X_n}(x_n) dx_1 \ldots dx_n = \frac{n}{n+1}(1 - \frac{1}{n})^{n-1}$

ii) the second term, is constant $1/n$: $\sum_{i=1}^{n} \int_0^1 \ldots \int_0^1 x_i \cdot \frac{1}{n} \cdot f_{X_1}(x_1) \cdot \ldots \cdot f_{X_n}(x_n) dx_1 \ldots dx_n = \sum_{i=1}^{n} \frac{1}{n} \cdot \frac{1}{2} = \frac{1}{2}$

iii) the thrid term, we integrate and add two cases separately, namely $i = k$ and $i \neq k$: $\sum_{i=1}^{n} \frac{1}{n} \sum_{k=1}^{n} \int_0^1 \ldots \int_0^1 x_i \cdot x_k^{n-1} \prod_{j \neq k}(1 - x_j^{n-1}) f_{X_1}(x_1) \cdot \ldots \cdot f_{X_n}(x_n) dx_1 \ldots dx_n =$

$\frac{1}{n+1}(1 - \frac{1}{n})^{n-1} + \frac{(n-1)^2}{2n(n+1)}(1 - \frac{1}{n})^{n-2}$

Combing the above results and simplifying give us the desired formula. ∎